# Final Cooling for a High-Energy High-Luminosity Lepton Collider

**David Neuffer,**[a*] **Hisham Sayed,**[b] **Terry Hart and Don Summers**[c]

[a] *Fermilab,*
  *PO Box 500, Batavia IL 60510, USA*
[b] *BNL,*
  *Upton, NY 11973, USA*
[c] *University of Mississippi*
  *Oxford, MS 38655, USA*

 *E-mail:* neuffer@fnal.gov

ABSTRACT: A high-energy muon collider scenario require a "final cooling" system that reduces transverse emittance by a factor of ~10 while allowing longitudinal emittance increase. The baseline approach has low-energy transverse cooling within high-field solenoids, with strong longitudinal heating. This approach and its recent simulation are discussed. Alternative approaches which more explicitly include emittance exchange are also presented. Round-to-flat beam transform, transverse slicing, and longitudinal bunch coalescence are possible components of an alternative approach. Wedge-based emittance exchange could provide much of the required transverse cooling with longitudinal heating. Li-lens and quadrupole focusing systems could also provide much of the required final cooling.

**Contents**

## 1. Introduction

The Muon Accelerator Program (MAP) has developed scenarios for future heavy-lepton (muon) colliders. An outline sketch of the scenario components is displayed in Figure 1 and potential parameters in Table 1.[1] Scenarios for a high-energy high-luminosity collider require cooling the beam transversely to ~0.00003m (rms, normalized emittance) while allowing a longitudinal emittance of ~0.1m (rms, normalized).[1] The present 6-D cooling systems cool the muons to ~0.0003m transversely and ~0.001m longitudinally.[2] Thus the collider scenarios require a "final cooling" system that reduces transverse emittances by a factor of ~10 while allowing longitudinal emittance increase. Previously, Palmer et al. have developed such a system, which includes transverse ionization cooling of low-energy muons within high field solenoids.[3, 4] At low-energies, the variation of momentum loss with energy anti-damps the beam longitudinally, increasing the longitudinal emittance, Figure 2 shows the progression of emittances throughout a collider cooling scenario, with the "final cooling" portion of that displayed as the lines with transverse emittance decrease with longitudinal emittance increase leading to final values at $\varepsilon_t$ = 25μ and $\varepsilon_L$ = ~30—60mm. More recently, Sayed et al. [5] have developed a detailed model of the final cooling system with G4Beamline tracking results that obtain performance similar to the Palmer baseline design. These systems and simulations are discussed below.

    Since this "final cooling" is predominantly an emittance exchange between transverse and longitudinal dimensions, it is possible that similar results could be obtained in a final cooling system that explicitly incorporates emittance exchanges, and avoid the very large magnetic

fields and very low-frequency rf with very-low-energy muons required at the end of the baseline systems. Approaches toward this are being developed by Summers et al. [6] A "round-to-flat" transform, much like that demonstrated at the Fermilab photoinjector,[7] could be used. This could be combined with a transverse beam slicer and longitudinal bunch recombiner, to obtain the small transverse emittance in both planes within a single bunch. This concept is described below, and variations which can reach the desired emittance goal with or without the round to flat transform are discussed.

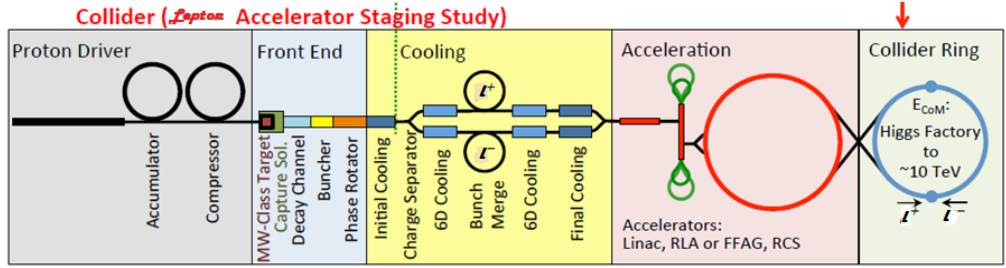

Table 1 Parameters of Muon Collider Scenarios

| Parameter | Unit | Higgs factory | 3 TeV design | 6 TeV design |
|---|---|---|---|---|
| Beam energy | TeV | 0.063 | 1.5 | 3.0 |
| Number of IPs |  | 1 | 2 | 2 |
| Circumference | m | 300 | 2767 | 6302 |
| $\beta^*$ | cm | 2.5 | 1 | 1 |
| Tune $v_x/v_y$ |  | 5.16/4.56 | 20.13/22.22 | 38.23/40.14 |
| Compaction | $1/\gamma_t^2$ | 0.08 | -2.88E-4 | -1.22E-3 |
| Emittance (Norm.) | mm·mrad | 300 | 25 | 25 |
| Momentum spread | % | 0.003 | 0.1 | 0.1 |
| Bunch length | cm | 5 | 1 | 1 |
| muons/bunch | $10^{12}$ | 2 | 2 | 2 |
| Repetition rate | Hz | 30 | 15 | 15 |
| Average luminosity | $10^{34}$ cm$^{-2}$s$^{-1}$ | 0.005 | 4.5 | 7.1 |

Large transfers in emittance between transverse and longitudinal dimensions can be obtained by passing focused beams through compact wedge absorbers. Transfers to very small transverse emittances are possible. Some parameters for such transfers are presented along with some concepts for integrating these exchanges into a final cooling design.

The initial approach to final cooling uses solenoid focusing for minimal emittance cooling. Li lens focusing and cooling can also obtain small emittances and can also be used in a final cooling system. Constraints and possibilities for Li-based systems are discussed.

Quadrupole based focusing can also obtain small-emittance focusing and will be needed in matching of asymmetric beams into and out of emittance reduction absorbers. Their use in a final cooling system is also discussed.

## 2. Baseline final cooling

### 2.1 Baseline scenario overview

A baseline approach to final cooling was developed by Palmer et al. [3,4]. The system is based on transverse ionization cooling of low-energy muons within high field solenoids, with lower energies and higher fields obtaining smaller $\varepsilon_t$. Within the cooling channel the transverse emittance decreases, with equilibrium values of:

$$\varepsilon_{N,eq} \cong \frac{\beta_t E_s^2}{2\beta mc^2 L_R (dE/ds)}$$

where $\varepsilon_{N,eq}$ is the equilibrium normalized transverse emittance in the cooling channel, $E_s = $ ~13.6MeV, $m = 105.66$ MeV/c$^2$ is the muon mass, $L_R$ and $dE/ds$ are the radiation length and energy loss rate in the cooling absorber material, and $\beta_t$ is the transverse betatron focusing function within the absorber. Within a solenoid of magnetic field B, this is:

$$\beta_t(m) \cong \frac{2P_\mu(GeV/c)}{0.3B(T)}$$

With $B$=40T and $p_{\mu}$=33 MeV/c ($E_\mu$ =5MeV), $\beta_t \approx 0.56$cm and $\varepsilon_{N,eq} \approx 0.00001$m.

The beam enters the final cooling system with the beam cooled by upstream 6-D cooling systems to minimal values of to $\varepsilon_t$ ~0.0003m transversely and $\varepsilon_L$ ~0.015m. For final cooling, the beam momentum is reduced initially to 135 MeV/c and only transverse cooling is used. The final cooling system consists of ~14 stages. Each stage consist of a high-field solenoid with an H$_2$ absorber within the magnet, followed by an rf and drift system within lower-field magnets to phase-rotate and reaccelerate the muons. From stage to stage, the muon beam energy is reduced (from 66 MeV toward 5MeV) and the magnet field strength is increased to minimize $\varepsilon_{N,eq}$.

At low energies, the energy loss is strongly antidamping (increasing the energy spread) and the longitudinal emittance increases dramatically, and the final cooling lattices do not include the emittance exchange needed to obtain longitudinal cooling. In the final stages of cooling, this antidamping is as large as the transverse damping; the 6-D emittance product $\varepsilon_t^2\varepsilon_L$ is roughly constant. In the final cooling scenario, the bunches are lengthened and rf rotated between absorbers to keep $dp/p < $ ~10%. This increases the bunch length from 5cm to $\sigma_{ct} = 4$m by the end of the system. The rf frequency decreases correspondingly, from ~201MHz at start to ~4MHz at the end. (RF frequencies < 20 MHz were considered unrealistic and the last five stages require induction linacs.)

Figure 2 graphically displays the parameters of the 14-stage system. The energy loss was simulated within ICOOL and the energy-phase motion tracked with a 1-D model, obtaining final emittances of $\varepsilon_t$ ~0.000025m transversely and $\varepsilon_L$ ~0.72m with ~33% beam loss in an ~76m long system.

Major challenges in the design include the cost of high-field magnets, the low-frequency rf, and the awkward deceleration and reacceleration of low-energy μ's. 50T and even 40T are somewhat above present capabilities, and require additional R&D as well as large construction and operation effects. Also, the baseline system is mostly emittance exchange between transverse and longitudinal degrees of freedom.

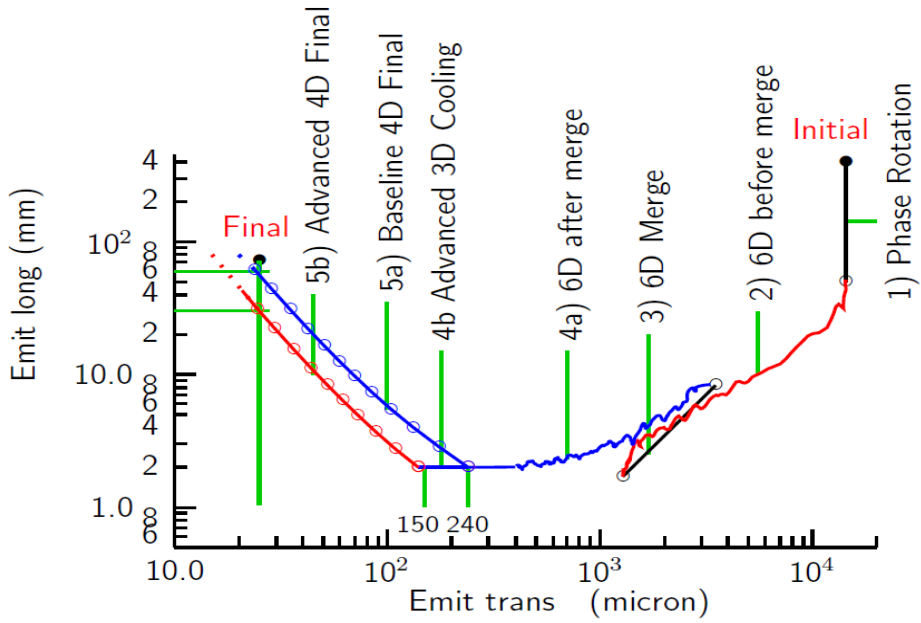

**Figure 1.** Overview of the evolution of emittance parameters for muon collider cooling systems.

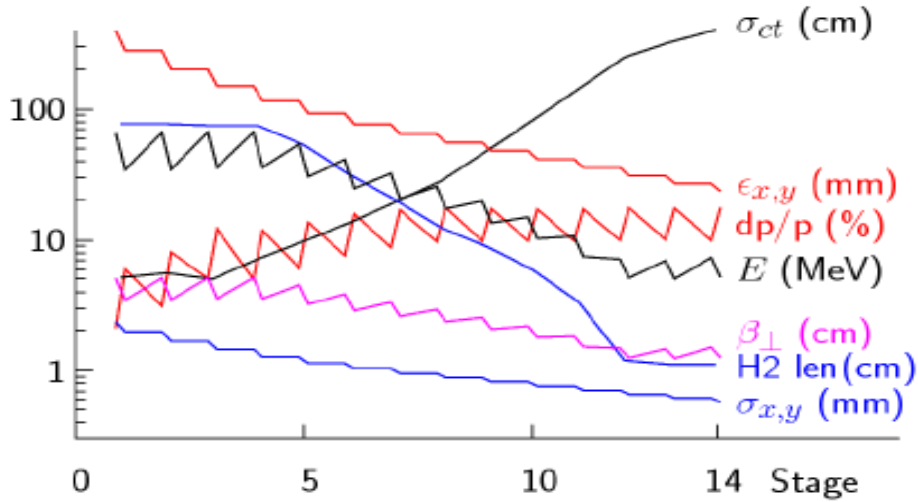

**Figure 2.** Overview of the evolution of beam and system parameters through a 14-stage final cooling system.

## 2.2 Simulation of a baseline scenario

The initial evaluations of the baseline scenario used a simplified model for simulation, and did not completely model the beam dynamics and matching from step to step. H. Sayed et al. have developed a more detailed and complete model of a final cooling channel and simulated it using the G4Beamline code.[8] In this first detailed simulation, the magnetic fields in the high-field magnets were limited to 25—32 T, and the cooling beam momenta ranged from 135 MeV/c to 70 MeV/c (40 to 20 MeV kinetic energy).

The channel consists of 16 stages. Each stage has five major components: a set of coaxial solenoidal coils providing the ~30T field for cooling, a liquid hydrogen absorber, two matching sections (before and after the high-field solenoid) with lower field solenoidal focusing (~3.5T) in the rest of the stage, with RF cavities for longitudinal phase-space rotation, and RF cavities for acceleration. Fig. 3 shows a stage of the system.

The beam momentum is reduced gradually along the channel from stage to stage to reduce the equilibrium emittance. ($\beta_t$ at the absorbers is reduced from 3.4 cm to 1.8 cm over the channel.) Each stage is separately tuned for acceptance and cooling. Field flips are introduced at 5 locations between stages to equalize the cooling of the transverse modes.

The disadvantage in low energy cooling is the increase in longitudinal emittance. At low energies, energy loss increases with reduced energy and this heats the beam longitudinally. The baseline channel does not incorporate any longitudinal cooling mechanisms, and must therefore accommodate this heating. This is done by lengthening the bunch in each stage, with phase-energy rf rotation that keeps the mean energy spread < ~3—4 MeV. The bunch length is initially $\sigma_{ct}$ =5 cm within 325 MHz rf and lengthens to $\sigma_{ct}$ =180 cm with 20 MHz rf.

The 16 stage channel (135m) was simulated using G4Beamline, which uses the GEANT4 physic libraries. Magnetic fields are computed using realistic coil and current settings. RF cavities are modelled as cylindrical pillboxes. A Gaussian input beam with $\varepsilon_{T,N}$ = 300 μ and $\varepsilon_L$ = 1.5mm was cooled to $\varepsilon_{T,N}$ = 55 μ and $\varepsilon_L$ = 1.5 mm, with a transmission (including decay) of 50%. Cooling performance is shown in figure 4. For a first attempt at a complicated problem with overall dynamic complexity, the calculated cooling performance is excellent.

The final transverse emittance is about a factor of 2 larger than the goal value and transmission could be improved. Further optimization and use of higher fields or lower momentum beams would be helpful and reach the desired goals.

The initial ~7 stages are relatively efficient in obtaining transverse cooling without large longitudinal increase, and beam losses.(<~10% beam loss) The following stages are progressively less efficient, with increasing losses (to 50%) and increasing longitudinal emittance. This suggests that an optimum final cooling system might incorporate the initial stages followed by a more efficient emittance exchange system replacing the final stages.

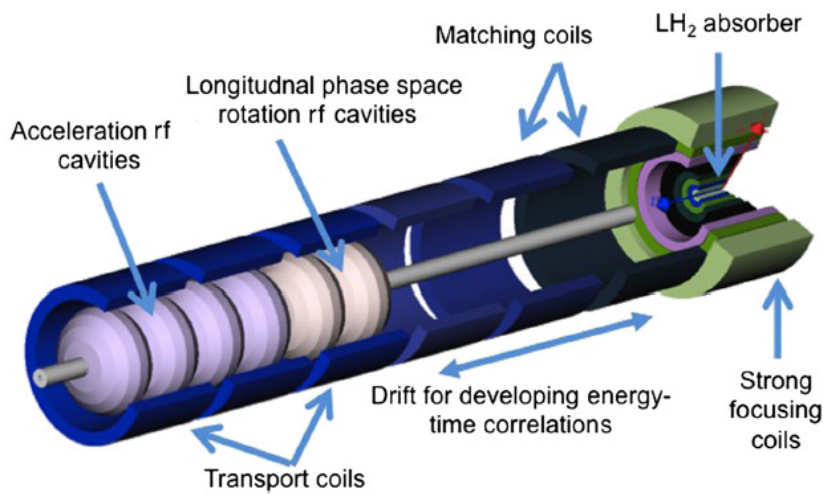

**Figure 3.** Overview of one final cooling stage. Each stage has strong coaxial focusing coils that enclose the LH$_2$ absorber folooed by matching cois, energy-pahse rotation Rf cavities and acceleration rf cavities.

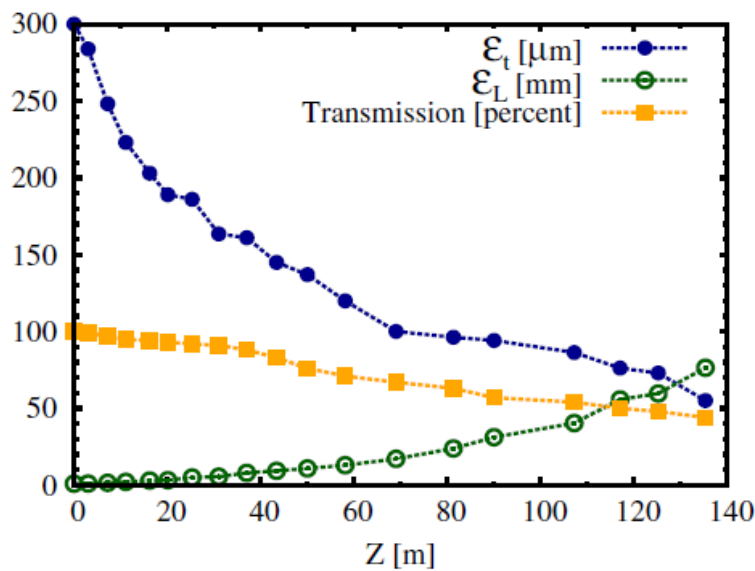

**Figure 4.** Transverse and longitudinal emittances along the 135m channel, along with transmission, which includes muon decay and dynamic losses.

### 2.3 Li lens based cooling

An alternative cooling method that can be particularly effective in cooling to minimal emittances is a Li lens –based cooling system. Strong focusing cooling is obtained by passing the beam through a conducting light-metal (Li) rod, which carries a pulsed electrical current. The rod is simultaneously a focusing element and an energy-loss absorber.[9, 10] The pulse provides an azimuthal magnetic field of:

$$B_\theta(r) = \frac{\mu_0 I r}{2\pi R_c^2},$$

where $R_c$ is the rod radius and $I$ is the total current in the rod. This provides radial focusing. The matched $\beta_t$ for beam at a momentum of $p_\mu$ in the rod is:

$$\beta_t = \left(\frac{B\rho}{B'_\theta}\right)^{1/2} = \left(\frac{p_\mu R_c}{e B_\theta(R_c)}\right)^{1/2}.$$

(With $p_\mu$=300MeV/c, $R_c$=2mm and $B_\theta(R_c)$ = 20T, $\beta_t$=1.0cm.)

Balbekov has designed a sequence of Li lens coolers with matching and rf for final cooling.[11] Matching was obtained by short high-field solenoids and 200—100 MHz rf was used. A sequence of 12 1m long Li lenses in a 120m long system reduces transverse emittances of ~250 MeV/c µ's from 400µ to 85µ with lens gradients increasing from 34 to 95 T/cm ($R_c$ decreases from 0.5 to 0.2cm). No longitudinal cooling was included. (Longitudinal emittance increased to ~10 mm.) Variations to reach 60—70µ transverse emittance, but with more losses and longitudinal dilution were discussed.

Li lens cooling was not included in the baseline because Li lens technology is not yet established at the level required. The high rep rate needed for a collider would probably require liquid-metal Li lenses. Also, final transverse emittances were somewhat above the high-energy collider goals. However, further emittance exchange in a multistage system (with wedges) could reach the desired transverse emittances.

## 3. Alternative front end scenarios

The baseline final cooling scenario is predominantly emittance exchange at extreme parameters. Therefore we may consider including other phase-space manipulations that obtain similar results.

### 3.1 Final Cooling with bunch slicing

An alternative approach to final cooling is presented by D. Summers et al.[6] Since the final cooling is dominated by emittance exchange, the approach here is to emphasize explicit emittance exchange and avoid the use of very-low frequency rf, very-low energy beams and high fields. The final cooling is envisioned as four stages:
1. Transverse Cooling. A cooling system similar to that of the baseline cooling system is used to cool the beam transversely within magnetic fields and rf systems that are relatively reasonable: $P_\mu$ = ~100MeV/c, B <30T, $f_{RF}$ >~150 MHz, without large beam loss. This would be much like the first 4—6 stages of the baseline system. Field-flips would not be placed between stages, enabling the development of cyclotron/drift asymmetry that can enable the round to flat transform. (A ratio of emittances > ~10 is possible.) The length of that system should be ~40m, and it should cool $\varepsilon_t$ to ~$10^{-4}$m, while $\varepsilon_L \rightarrow$ ~0.004m.
2. Round to flat beam transform. Following the technique developed for the ILC injector and other applications,[7] a solenoid $\rightarrow$ three skew-quad system transforms a "round" (large drift, small cyclotron modes)to a flat (large $x$, small $y$) emittance: $\varepsilon_t \rightarrow \varepsilon_x$ = 0.0004, $\varepsilon_y$ = 0.000025. (These particular numbers assume a drift/cyclotron ratio of 16.)
3. Transverse slicing. The beam is sliced using multiple passes through "slow-extraction –like" septa into a string of bunches (~16). The slices are in the larger emittance transverse plane, obtaining bunches with equal transverse emittances: $\varepsilon_x$ = 0.000025, $\varepsilon_y$ = 0.000025.
4. Longitudinal recombination. The train of ~16 bunches is accelerated to a larger energy (~10 GeV?), where a snap coalescence in a medium-energy storage ring combines these into a single bunch with enlarged longitudinal emittance ($\varepsilon_x$ = 0.000025, $\varepsilon_y$ = 0.000025, $\varepsilon_L$ =~ 0.064m).

Similar manipulations are possible without use of the "round to flat" process. The sequence could be:
1. Transverse Cooling. A cooling system to minimize emittances within reasonable fields is used, as in the step 1 of the above scenario. It should cool $\varepsilon_x$ and $\varepsilon_y$ to ~$10^{-4}$m, while $\varepsilon_L \rightarrow$ ~0.004m or less.
2. Transverse slicing. The beam is sliced using multiple passes through a "slow-extraction–like" thin septum into a string of bunches (~10). The slices are in one plane, obtaining bunches with asymmetric emittances: $\varepsilon_x$ = 10μ, $\varepsilon_y$ = 100μ.
3. Longitudinal recombination. The bunches are accelerated into a ring that combines them into a single bunch ($\varepsilon_x$ = 10μ, $\varepsilon_y$ = 100μ, $\varepsilon_L$ =~ 0.04m).
4. The beams accelerate and collide as flat beams, Collisions of $\varepsilon_x$ = 10μ, $\varepsilon_y$ = 100μ could be matched in luminosity to $\varepsilon_{t\,=\,(\varepsilon_x\,\varepsilon_y)}{}^{1/2}$ =~30μ round beams.

Flat beam collisions have some advantages. Chromaticity correction is much easier, and detector shielding could be simpler. However, luminosity may be decreased by the "hour glass" effect, if $\beta_x^* \ll$ bunch length.

### 3.2 Comments on round/flat beams

Many ionization cooling systems are dominated by solenoidal focusing. Within solenoidal fields, the eigenmodes (+ and -) are associated with drift (d) and cyclotron (k) modes, respectively; x and y coordinates are not eigenmodes.[12, 13]  The k mode coordinates are:

$$\begin{pmatrix} \kappa_1 \\ \kappa_2 \end{pmatrix} = \sqrt{\frac{c}{eB}} \begin{pmatrix} k_y \\ k_x \end{pmatrix} = \sqrt{\frac{c}{eB}} \begin{pmatrix} p_y + \frac{eB}{2c} x \\ p_x - \frac{eB}{2c} y \end{pmatrix}$$

and are simply proportional to the kinetic momentum coordinates $k_x$, $k_y$. The d coordinates are:

$$\begin{pmatrix} \xi_1 \\ \xi_2 \end{pmatrix} = \sqrt{\frac{eB}{c}} \begin{pmatrix} d_x \\ d_y \end{pmatrix} = \sqrt{\frac{eB}{c}} \begin{pmatrix} x - \frac{c}{eB} k_y \\ y + \frac{c}{eB} k_x \end{pmatrix} = \sqrt{\frac{eB}{c}} \begin{pmatrix} \frac{x}{2} - \frac{c}{eB} p_y \\ \frac{y}{2} + \frac{c}{eB} p_x \end{pmatrix}$$

and are proportional to the centers of the Larmor motion, associated with the position coordinates. Within a constant B field the k mode is damped, while the d mode is not. Field flips exchange k and d modes, and can balance the emittance damping. Most previously designed cooling systems use frequent field flips and therefore obtain round beams with equal emittances in x,y or k, d coordinates.

Without field flips, solenoidal cooling can develop a large emittance asymmetry between modes. The 4-D emittance is

$$\varepsilon_{4D} = \varepsilon_T^2 = \varepsilon_+ \varepsilon_- = (\varepsilon_P + L)(\varepsilon_P - L)$$

where $2L$ is the angular momentum and $\varepsilon_P$ is the projected emittance. Edwards et al.[14] have shown that a skew quad transport can translate $\varepsilon_+$ and $\varepsilon_-$ into $\varepsilon_x$ and $\varepsilon_y$ (decoupled), and vice versa. If $\varepsilon_+$ and $\varepsilon_-$ are very different, a "round" beam is transformed to a "flat" beam outside the solenoids. The process has been demonstrated in $e^-$ beams. Cooling of muon beams to $\varepsilon_+/\varepsilon_- \gg 10$, with $\varepsilon_-$ smaller than in a symmetric system, has also been simulated.

The use of some non-flip cooling to develop asymmetric beams coupled with round to flat transforms adds flexibility in the generation of higher luminosity configurations at the end of final cooling, and can be incorporated into final cooling scenarios. We have suggested some in the present discussion, but other and better variations may be developed in the future.

## 4. Thick Wedge emittance exchange

Much of the final cooling is an emittance exchange, with longitudinal heating nearly equal to the transverse cooling. The simplest form of emittance exchange is found by passing the beam through a wedge absorber, where the bunch width is transformed into an energy width. It was previously noted that large emittance exchanges by single wedges are possible near final cooling parameters.[15] A formalism for estimating exchanges obtainable from single wedges was presented. A reevaluation using the present beam parameters shows that larger exchanges than initially suggested are possible.

Figure 5 shows a stylized view of the passage of a beam with dispersion $\eta_0$ through an absorber. The wedge is approximated as an object that changes particle momentum offset $\delta = \Delta p/P_0$ as a function of $x$, and the wedge is shaped such that that change is linear in $x$. (The change in average momentum $P_0$ is ignored, in this approximation. Energy straggling and multiple scattering are also ignored.) The rms beam properties entering the wedge are given by the transverse emittance $\varepsilon_0$, betatron amplitude $\beta_0$, dispersion $\eta_0$ and relative momentum width $\delta_0$. (To simplify discussion the beam is focussed to a betatron and dispersion waist at the wedge: $\beta_0'$, $\eta_0' = 0$. This avoids the complication of changes in $\beta'$, $\eta'$ in the wedge.) The wedge is represented by its relative effect on the momentum offsets $\delta$ of particles within the bunch at position $x$:

$$\frac{\Delta p}{p} = \delta \rightarrow \delta - \frac{(dp/ds)\tan\theta}{P_0} x = \delta - \delta' x$$

$dp/ds$ is the momentum loss rate in the material ($dp/ds = \beta^{-1}dE/ds$). $x\tan\theta$ is the wedge thickness at transverse position $x$ (relative to the central orbit at $x=0$), and $\delta' = dp/ds \tan\theta /P_0$ to indicate the change of $\delta$ with $x$.

Under these approximations, the initial dispersion and the wedge can be represented as linear transformations in the $x$-$\delta$ phase space projections and the transformations are phase-space preserving. the dispersion can be represented by the matrix:

$$\mathbf{M}_\eta = \begin{bmatrix} 1 & \eta_0 \\ 0 & 1 \end{bmatrix}$$, since $x \Rightarrow x + \eta_0 \delta$. The wedge can be represented by the matrix:

$$\mathbf{M}_\delta = \begin{bmatrix} 1 & 0 \\ -\delta' & 1 \end{bmatrix}$$, so that the dispersion + wedge becomes: $\mathbf{M}_{\eta\delta} = \begin{bmatrix} 1 & \eta_0 \\ -\delta' & 1-\delta'\eta_0 \end{bmatrix}$. This

matrix changes the dispersion, the momentum width, and the transverse beam size (dispersion removed). Writing the $x$-$\delta$ beam distribution as a phase-space ellipse: $g_0 x^2 + b_0 \delta^2 = \sigma_0 \delta_0$, and transforming the ellipse by standard betatron function transport techniques (obtaining coefficients $b_1$, $g_1$, $a_1$),[16] we obtain new beam parameters. The momentum width is changed to:

$$\delta_1 = \sqrt{g_1 \sigma_0 \delta_0} = \delta_0 \left[ (1-\eta_0\delta')^2 + \frac{\delta'^2 \sigma_0^2}{\delta_0^2} \right]^{1/2}.$$

The bunch length is unchanged. The longitudinal emittance has therefore changed simply by the ratio of energy-widths, which means that the longitudinal emittance has changed by the factor $\delta_1/\delta_0$. The transverse emittance has changed by the inverse of this factor:

$$\varepsilon_1 = \varepsilon_0 \left[ (1-\eta_0\delta')^2 + \frac{\delta'^2 \sigma_0^2}{\delta_0^2} \right]^{-1/2}.$$ The new values of $(\eta, \beta)$ are:

$$\eta_1 = -\frac{a_1}{g_1} = \frac{\eta_0(1-\eta_0\delta') - \delta' \frac{\sigma_0^2}{\delta_0^2}}{(1-\eta_0\delta')^2 + \delta'^2 \frac{\sigma_0^2}{\delta_0^2}}, \text{ and } \beta_1 = \beta_0 \left[ (1-\eta_0\delta')^2 + \frac{\delta'^2 \sigma_0^2}{\delta_0^2} \right]^{-1/2}.$$

Note that the change in betatron functions ($\beta_1$, $\eta_1$) implies that the following optics should be correspondingly rematched.

As currently presented, the wedge exchanges emittance between one transverse dimension and longitudinal; the other transverse plane is unaffected. Serial wedges could be used to balance $x$ and $y$ exchanges, or a more complicated coupled geometry could be developed.

Wedge parameters can be arranged to obtain large exchange factors in a single wedge. In upstream systems the wedges can be arranged to obtain a factor of longitudinal cooling (at expense of transverse heating). In final cooling we wish to reduce transverse emittance at the cost of increased longitudinal emittance.

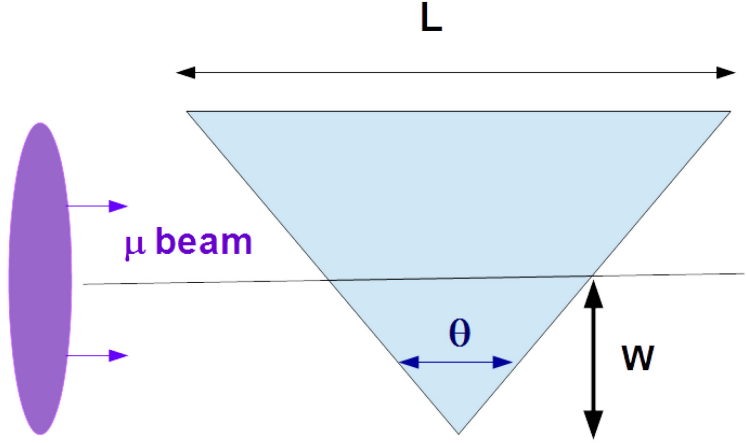

**Figure 5.** Schematic view of a muon beam passing through a wedge.

### 4.1 Thick wedges at final cooling parameters

For final cooling, the beam and wedges should be matched to obtain a large factor of increase in momentum spread. That means that the energy spread induced by the wedge should be much greater than the initial momentum spread: $\delta_0 \ll \delta'\sigma_0 = \dfrac{2\tan\left(\frac{\theta}{2}\right)\frac{dp}{ds}}{P_0}\sigma_0$. Thus the incident beam should have a small momentum spread and small momentum $P_0$ and the wedge should have a large $\tan(\theta/2)$, large $dp/ds$ and a large $\sigma_0 = (\varepsilon_0\beta_0)^{1/2}$. ($\varepsilon_0$ is unnormalized, rms in this section.) Beam from a final cooling segment (high-field solenoid or Li lens) is likely to have $P_0 \approx 100$—150 MeV/c, and $\delta p \approx 3$ MeV/c. For optimum single wedge usage, $\delta p$ should be reduced to ~0.5MeV/c, and this can be done by rf debunching of the beam to a longer bunch length.

For maximal wedge effect, the beam size should be matched to the wedge size $w$ ($w \approx 2\sigma_0$). To minimize multiple scattering heating at the wedge, $\beta_0$ should be small (< a few cm) and the wedge should be a low-Z material. From

$$\frac{dp}{ds} = \frac{1}{\beta c}4\pi N_A r_e^2 m_e c^2 \rho \frac{Z}{A}\left[\frac{1}{\beta^2}\ln\left(\frac{2m_e c^2 \gamma^2 \beta^2}{I(Z)}\right) - 1 - \frac{\delta}{2\beta^2}\right],$$ the material should have a

high density $\rho$ to obtain large $dp/ds$. $dp/ds$ increases as $p$ decreases, which implies using smaller $P_0$. (This variation should also be considered in modifying the wedge shape for small $P_0$; an uncorrected linear wedge enlarges the momentum spread, increasing the longitudinal emittance.) Materials under consideration are Be, C (graphite or diamond density), BeO, …; diamond-density C is preferred but solutions with Be, lower-density C and BeO are possible.

At final cooling parameters ($\beta_0 \approx 1$cm, $\varepsilon_0 \approx 100\mu$), $\sigma_0 \approx 1$mm, which means the final cooling wedges would be only a few mm in size. However at final cooling parameters large exchanges are indeed possible.

The effect has been simulated using the simulation code ICOOL,[17] initially using linear wedges. For a particular example we consider a high-density C wedge (diamond density) with an input μ beam at 100 MeV/c, $\delta E = 0.5$ MeV ($\delta p = 0.73$ MeV/c), $\varepsilon_x = \varepsilon_y = 0.013$cm, matched to $\beta_t$

= 2cm at the center of the wedge ($\sigma_x$=1.65mm), and zero initial dispersion. The wedge parameters are: $w$=3mm, $\theta$=85° (~5.6mm thick at beam center).

Simulation results are presented in Table 2 and figure 6, with emittances calculated using EMITCALC (part of the ICOOL tools). The wedge-plane (x) emittance is reduced to 0.0025cm (a factor of 5.2 !) while the y-plane emittance is unaffected. However, the energy spread is increased to ~3.9MeV, diluting 6-D emittance by ~50%. Some of this dilution may be correctable by improving the phase-space match and using higher order correction. In any case, this simple wedge does succeed in getting transverse emittance below the goals of a high-energy collider.

Table 2: Beam parameters at entrance, center and exit of a w=3mm, θ=85° diamond wedge. The $z$ = 0, 0.6, 1.2cm rows are beam parameters before, at the center, and after the wedge. The 0.6cm values also indicate results that can be obtained with a half-strength wedge.

| $z$(cm) | $P_z$(MeV/c) | $\varepsilon_x(\mu)$ | $\varepsilon_y(\mu)$ | $\varepsilon_L$ (mm) | $\sigma_E$(MeV) | 6-D $\varepsilon$ increase |
|---|---|---|---|---|---|---|
| 0 | 100 | **129** | 127 | 1.0 | 0.50 | 1.0 |
| 0.6 | 95.2 | **40.4** | 130 | 4.03 | 1.95 | 1.29 |
| 1.2 | 90.0 | **25.0** | 127 | 7.9 | 3.87 | 1.54 |

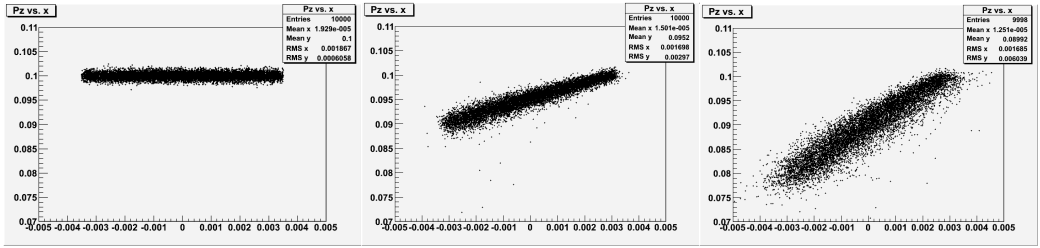

**Figure 6**. $x$-$P_z$ distributions for μ beam passing through a wedge system, shown at entrance, middle and end of the wedge.

The beam from a wedge transform could be reaccelerated and phase-energy rotated to a longer bunch with small $\delta E$ for a pass through a second wedge. If properly rematched, ICOOL simiulations indicate that the second wedge could reduce vertical emittance to ~0.0025mm, while horizontal emittance increase could be limited to keep $\varepsilon_x$ < 0.03mm (30μ).

With this addition a final cooling scenario using as few as 2 wedges can be considered. The sequence could be:

1. Transverse Cooling. A cooling system to minimize emittances within reasonable fields is used, as in the step 1 of the above scenario. It should cool $\varepsilon_x$ and $\varepsilon_y$ to ~1.3×10$^{-4}$m, while $\varepsilon_L$≈~0.003m. This could be the initial sector of the baseline front end.

2. Match into first wedge: The beam is stretched to $\sigma_z$ = ~0.6m to enable phase energy rotation to $\delta E$ < 0.5 MeV while being decelerated to ~100 MeV/c. Focus onto the first wedge causes an emittance exchange to $\varepsilon_x$ = 25μ, $\varepsilon_y$ = 130μ, $\varepsilon_L$ =~0.015m .

3. Match into second wedge: The beam is stretched to $\sigma_z$= ~3m to enable phase energy rotation to δE < 0.5 MeV while being accelerated to ~100 MeV/c. Focus onto the second wedge places $\varepsilon_x$ = 30μ, $\varepsilon_y$ = 25μ, $\varepsilon_L$ =~0.075m. .

4. The beam is phase-energy rotated and accelerated and bunched in a 12m long bunch train (12 bunches at 300 MHz or 24 at 600 MHz).

5. Longitudinal recombination. The bunches are accelerated into a ring that combines them by snap coalescence into a single bunch ($\varepsilon_x < 30\mu$, $\varepsilon_y < 30\mu$, $\varepsilon_L =\sim 0.075$m).

In this outline scenario, phase space dilution is limited to ~25% in each wedge. The beam exiting the wedge has a transverse $\beta_t$ of ~0.5cm and dispersion $\eta$ of ~5cm, and the downstream optics must adequately match through the transport without excessive emittance dilution. A detailed design has not yet been developed and would be an important subject for future research and optimization.

The scenario above maximizes the amount of exchange in single wedges. An alternate strategy would be the split up the exchanges into a series of shorter wedges. While that would increase the number of wedge to cooler/rf transition regions, the transitions could be more efficient, reducing emittance dilution. (Smaller energy loss per wedge may or may not improve overall matching.) The gains possible from increasing the number of wedges should be explored.

At the present level of analysis, the wedge exchange compares favourably with the last ~10 stages of the baseline final cooling system, obtaining transverse emittances within the high-energy collider goals, without greater longitudinal dilution, and without large beam losses from decay and aperture loss. However, the process has not been simulated and optimized in as much detail and is not yet integrated into a full scenario. Nonetheless, the initial evaluations are so promising that it appears that some amount of wedge exchange should be included in a final optimized system.

## 4.2 Thick wedge experiment in MICE

The MICE experiment has considered inserting a wedge absorber into the beam line for measurements of emittance exchange cooling.[18] The MICE experiment has considerable flexibility in beam definition, and can obtain an initial beam with small δp by selecting particle tracks from the ensemble measured in MICE within that momentum band. Passage of that beam through a wedge will cause emittance exchange and large exchange factors can be obtained with matched parameters. The intrinsic beam sizes are larger ($\varepsilon_{t,n} = $ ~0.003m, $\beta_t = \sim 0.4$ m) than the final cooling cases, and can be matched to larger absorbers. The resulting parameters would be a larger scale model of final cooling exchanges, and the experiment would be a direct demonstration of the underlying principle.

As an example we consider using a polyethylene ($CH_2$) absorber with $w$=5cm, $\theta$=60°, with the wedge oriented along $x$. (A Be or LiH wedge would have superior performance, but greater expense, and would not greatly improve the initial proof of principle demonstration.) With the incident beam matched to $\sigma_x$ = 2.5 cm, $P_0$=200 MeV/c and $\delta p$ = 2 MeV/c, one obtains an increase in $\delta p$ by a factor of ~4 accompanied by a reduction in $\varepsilon_x$ by a factor of ~4. This example was simulated in ICOOL, with results presented in table 3 and displayed in Fig. 7. The resulting scenario would be an interesting scaled model of a final cooling scenario and would test the basic physics and optics of the exchange configuration.

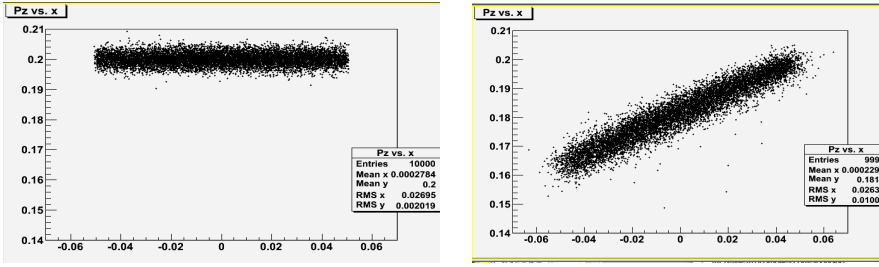 The $z$ = 0, 6, 12cm rows are beam parameters before, at the center, and after the wedge. The 6cm values also indicate results that can be obtained with a half-strength wedge.

**Figure 7.** $x$-$P_z$ projections of the beam before and after the polyethylene wedge. The $x$-axis shows -0.06 to +0.06m bins and $y$ axis shows $P_z$ from 0.14 to 0.21 GeV/c.

**Table 3:** Results of ICOOL simulation of μ beam passing through a wedge (CH$_2$, $w$=5cm, $\theta$=60°)

| $z$(cm) | $P_z$ | $\varepsilon_x$(mm) | $\varepsilon_y$ | $\varepsilon_L$(mm) | $\sigma_E$(MeV) | 6-D $\varepsilon$ increase |
|---|---|---|---|---|---|---|
| 0 | 200 | 4.32 | 4.25 | 3.125 | 1.95 | 1.0 |
| 6 | 192.9 | 1.94 | 4.22 | 7.41 | 3.91 | 1.06 |
| 12 | 181.9 | 1.06 | 4.11 | 14.6 | 8.68 | 1.11 |

### 4.3 Other cooling scenarios

Other cooling concepts that may improve final cooling parameters are under development. Y. Derbenev et al.[19] are considering a resonance based cooling scheme (parametric resonance ionization cooling) to obtain very small $\beta_t$ at absorbers. These would be combined with wedge-based exchanges, similar to those described above, but within resonance focusing. The resonance balances beam dynamics at the edge of instability, with the damping process insuring stability, while the resonance forces the beam into reduced transverse beam size.

Acosta, et al. [20] are considering quadrupole-based focusing to obtain $\beta_t$ =~1cm periodic systems, to obtain $\varepsilon_t$ = ~10$^{-4}$m while keeping $\varepsilon_L$ = ~0.02m. If successfully developed, the cooling system could be the initial part of the final cooling system, to be followed by beam slicing or wedge exchange sections.

When reliable and effective cooling is demonstrated in full simulations, these methods can be integrated into improved final cooling scenarios, perhaps permitting luminosities beyond the initial goals presented in table 1.

### 5. Summary and Conclusions

Final cooling for a high luminosity lepton (muon) collider requires substantial transverse emittance reduction. A baseline method using very high field solenoids with low energy beam has been developed and simulated. There are many possible paths toward better cooling and/or lower cost and/or better performance in obtaining high luminosity collisions, and some of these have been presented. Many of these additional concepts should be included in the system. Research is needed to produce an optimal final cooling system; performance beyond initial design goals is probable.

**Acknowledgments**

We thank the MAP collaboration for support and assistance in this research. Fermilab is operated by Fermi Research Alliance, LLC under Contract No. De-AC02-07CH11359 with the U. S. Department of Energy.